\documentclass[twocolumn,showpacs,superscriptaddress,preprintnumbers,prd]{revtex4-1}
\usepackage{graphicx,amsmath,amssymb,dcolumn,bm}
\usepackage{epic,eepic}
\usepackage[dvipdfmx]{pict2e}

\begin{document}
\preprint{KEK-TH-1906, J-PARC-TH-0053}
\title{Theoretical estimate on tensor-polarization asymmetry \\
in proton-deuteron Drell-Yan process}
\author{S. Kumano}
\affiliation{KEK Theory Center,
             Institute of Particle and Nuclear Studies, \\
             High Energy Accelerator Research Organization (KEK), \\
             1-1, Ooho, Tsukuba, Ibaraki, 305-0801, Japan}
\affiliation{J-PARC Branch, KEK Theory Center,
             Institute of Particle and Nuclear Studies, KEK, \\
           and
           Theory Group, Particle and Nuclear Physics Division, 
           J-PARC Center, \\
           203-1, Shirakata, Tokai, Ibaraki, 319-1106, Japan}
\affiliation{Department of Particle and Nuclear Physics, \\
             Graduate University for Advanced Studies (SOKENDAI), \\
             1-1, Ooho, Tsukuba, Ibaraki, 305-0801, Japan}
\author{Qin-Tao Song}
\affiliation{KEK Theory Center,
             Institute of Particle and Nuclear Studies, \\
             High Energy Accelerator Research Organization (KEK), \\
             1-1, Ooho, Tsukuba, Ibaraki, 305-0801, Japan}
\affiliation{Department of Particle and Nuclear Physics, \\
             Graduate University for Advanced Studies (SOKENDAI), \\
             1-1, Ooho, Tsukuba, Ibaraki, 305-0801, Japan}
\date{August 14, 2016}
\begin{abstract}
Tensor-polarized parton distribution functions are new quantities
in spin-one hadrons such as the deuteron, and they 
could probe new quark-gluon dynamics in hadron and nuclear physics. 
In charged-lepton deep inelastic scattering (DIS), they are
studied by the twist-two structure functions $b_1$ and $b_2$.
The HERMES collaboration found unexpectedly large $b_1$ values than
a naive theoretical expectation based on the standard deuteron model.
The situation should be significantly improved in the near future 
by an approved experiment to measure $b_1$ at JLab (Thomas Jefferson 
National Accelerator Facility). 
There is also an interesting indication in the HERMES result that 
finite antiquark tensor polarization exists. It could play an important role
in solving a mechanism on tensor structure in the quark-gluon level. 
The tensor-polarized antiquark distributions
are not easily determined from the charged-lepton DIS; however,
they can be measured in a proton-deuteron Drell-Yan process with
a tensor-polarized deuteron target. In this article, we estimate
the tensor-polarization asymmetry for a possible Fermilab Main-Injector 
experiment by using optimum tensor-polarized PDFs to explain the HERMES 
measurement. We find that the asymmetry is typically a few percent. 
If it is measured, it could probe new hadron physics,
and such studies could create an interesting field of high-energy 
spin physics. In addition, we find that a significant tensor-polarized 
gluon distribution should exist due to $Q^2$ evolution, 
even if it were zero at a low $Q^2$ scale. 
The tensor-polarized gluon distribution has never been observed, 
so that it is an interesting future project.
\end{abstract}
\maketitle

\section{Introduction}\label{intro}

It was discovered by the European Muon Collaboration (EMC) collaboration
in the measurement of the polarized structure function $g_1$
that only a small fraction of nucleon spin is carried by quarks
\cite{nucleon-spin}, which is in contradiction to the naive quark model. 
Since then, many theoretical and experimental efforts have been made
to clarify the origin of the nucleon spin. There could be contributions 
from gluon spin and partonic angular momenta. Theoretical and experimental 
efforts are in progress to solve the issue.

On the other hand, there are new polarized structure functions
\cite{fs83,hjm89}, which do not exist in the spin-1/2 nucleon, 
for spin-one hadrons and nuclei such as the deuteron. 
In charged-lepton deep inelastic scattering, they are named as 
$b_1$, $b_2$, $b_3$, and $b_4$ \cite{hjm89}. 
Projection operators of $b_{1-4}$ on the hadron tensor $W_{\mu\nu}$ 
are obtained in Ref. \cite{kk08}. The twist-two functions
are $b_1$ and $b_2$, and they are related with each other by
the Callan-Gross like relation $2x b_1 = b_2$ in the Bjorken
scaling limit. Therefore, it is interesting to investigate 
the leading-twist one $b_1$ (or $b_2$) first. 
A useful sum rule for the twist-two function $b_1$ was
proposed in Ref. \cite{b1-sum} by using the parton model,
and it could be used as a guideline for the existence
of tensor-polarized antiquark distributions.
On the other hand, proton-deuteron Drell-Yan processes are 
theoretically formulated in Ref. \cite{pd-drell-yan} for
the polarized deuteron including tensor polarization.

The leading-twist structure function $b_1$ probes a peculiar
aspect of internal structure in a spin-one hadron. It vanishes
if internal constituents are in the S wave, which indicates 
that it is a suitable observable for probing a dynamical aspect, 
possibly an exotic one, inside the hadron. In fact, the first measurement
of $b_1$ by the HERMES collaboration \cite{hermes05} indicated 
that the magnitude of $b_1$ is much larger than the one expected
by the standard deuteron model with D-state admixture
\cite{hjm89,kh91}. There could be other effects from
pions \cite{miller-b1} and shadowing phenomena 
\cite{b1-shadowing} in the deuteron.
There is also a suggestion that $b_1$ studies could lead 
to a new finding on a hidden-color component \cite{miller-b1}.
On related spin-one hadron physics, there are investigations on 
leptoproduction of spin-one hadron \cite{rho-production},
fragmentation functions \cite{spin-1-frag}, 
generalized parton distributions \cite{spin-1-gpd}, 
target-mass corrections \cite{mass-corr},
positivity constraints \cite{dmitrasinovic-96},
lattice QCD estimate \cite{lattice},
and angular momenta for spin-1 hadron \cite{angular-spin-1}.

The first measurement of $b_1$ was done by the HERMES collaboration 
in 2005 \cite{hermes05}. Its data indicated that $b_1$ has 
interesting oscillatory behavior as the function of $x$
and that the magnitude of $x b_1$ is of the order of $10^{-3}$.
Since $b_1$ is expressed by tensor-polarized
parton distribution functions, possible quark and antiquark 
distributions ($\delta_T q(x)$, $\delta_T \bar q(x)$) were extracted
from the HERMES data \cite{tensor-pdfs}. The analysis suggested
finite tensor-polarized antiquark distributions from the data
at $x<0.1$. Although the sign change of $b_1$ is expected
in a convolution description for the deuteron with D-state admixture,
the HERMES data indicate much larger $| \, b_1 |$. Therefore, 
the $b_1$ is expected to probe non-conventional physics beyond 
the standard deuteron model.
The HERMES measurement also suggested a new phenomena on
a finite tensor polarization for antiquarks.
There is a sum rule for $b_1$ \cite{b1-sum}, and a deviation
from this sum indicated the finite tensor-polarized
antiquark distributions in the similar way to Gottfried 
sum-rule violation \cite{flavor3}.

The deuteron tensor structure has been investigated for a long time
at low energies. However, time has come to investigate it,
through the tensor-polarized structure functions
and parton distribution functions, in terms of quark and gluon degrees 
of freedom. Furthermore, these quantities could be sensitive to exotic 
features such as the hidden color \cite{miller-b1}. 
In the HERMES measurement, there is already a hint that new hadron physics 
is needed to interpret its data.
Therefore, a new field of high-energy spin physics could be created
by investigating the tensor structure functions, as the EMC measurement
created the field of high-energy spin physics for the spin-1/2 nucleon.
The current situation on $b_1$ and tensor-polarized PDFs is summarized
in Ref. \cite{sk14}.

By considering this prominent prospect, the 
Thomas Jefferson National Accelerator Facility (JLab) experiment 
was approved for measuring the structure function $b_1$ \cite{Jlab-b1},
and the actual experiment will start in a few years.
In addition, the related tensor polarization $A_{zz}$ can be 
investigated at JLab in the large-$x$ region \cite{azz}.
These $b_1$ and $A_{zz}$ could be also investigated at the future 
Electron-Ion Collider (EIC) \cite{eic}.

In a simple model for the deuteron, it is not obvious to have 
a tensor-polarized antiquark distribution; however, a finite 
value is indicated in the HERMES experiment for the antiquark
tensor polarization. The best way to probe 
the antiquark distributions is to use a Drell-Yan process 
with a tensor-polarized deuteron target \cite{pd-drell-yan}. 
The Drell-Yan process for unpolarized proton - tensor-polarized deuteron 
is possible, and its measurement is now under consideration at Fermilab
\cite{Fermilab-MI}. However, there is no theoretical estimate 
on the tensor-polarized spin asymmetry, so that it is necessary 
to show even the order of magnitude for an experimental proposal,
especially for considering beam-time allocation in an actual measurement.
The purpose of this article is to show expected
spin asymmetries for the Fermilab measurement.

In this article, we explain formalisms first in Sec.\,\ref{tensor-pdfs},
especially on the tensor-polarized structure functions
and parton distribution functions (PDFs) in Sec.\,\ref{b1}, and 
the tensor-polarized Drell-Yan spin asymmetry is expressed
in terms of the tensor-polarized PDFs in Sec.\,\ref{pd-drell-yan}.
Then, our estimates on the spin asymmetry are shown for the possible
Fermilab experiment in Sec.\,\ref{results}.
The results are summarized in Sec.\,\ref{summary}.

\section{Tensor-polarized distribution functions for spin-one deuteron}
\label{tensor-pdfs}

We explain basic formalisms involving tensor-polarized structure functions
and PDFs in deep inelastic charged-lepton scattering and proton-deuteron
Drell-Yan process.

\subsection{Tensor-polarized structure functions \\
in charged-lepton deep inelastic scattering}
\label{b1}

First, the charged-lepton deep inelastic scattering (DIS) 
from the polarized deuteron is explained. The polarized DIS formalism
for the charged-lepton from the nucleon is well known, and it is 
generally expressed in terms of four structure functions, $F_1$, $F_2$,
$g_1$, and $g_2$. In addition to these functions, there exist four new
structure functions, $b_1$, $b_2$, $b_3$, and $b_4$, in the DIS from
the spin-one hadron such as the deuteron. 

\begin{figure}[t]
   \vspace{-0.30cm}
\begin{center}
   \includegraphics[width=4.5cm]{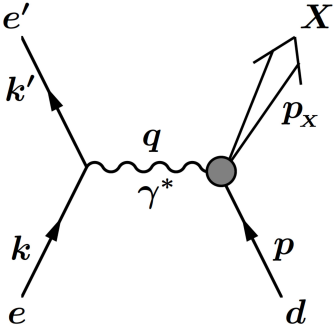}
\end{center}
\vspace{-0.5cm}
\caption{Deep inelastic scattering of charged lepton 
($e$) from spin-one hadron $d$.}
\label{fig:charged-lepton-dis}
\end{figure}

\begin{widetext}
In the charged-lepton DIS shown in Fig. \ref{fig:charged-lepton-dis},
the hadron tensor $W_{\mu \nu}$ is generally expressed 
for a spin-one hadron as \cite{hjm89,kk08,sk14}
\begin{align}
W_{\mu \nu}^{\lambda_f \lambda_i}
 = &  \frac{1}{4 \pi M} 
               \int d^4 \xi \, e^{i q \cdot \xi} \,
               \langle \, p, \lambda_f \, | \, [ \, J_\mu^{\, em} (\xi) ,
                      J_\nu^{\, em} (0) ]  \, | \, p, \lambda_i \, \rangle
\nonumber \\
   = & -F_1 \hat{g}_{\mu \nu} 
     +\frac{F_2}{M \nu} \hat{p}_\mu \hat{p}_\nu 
     + \frac{ig_1}{\nu}\epsilon_{\mu \nu \lambda \sigma} q^\lambda s^\sigma  
     +\frac{i g_2}{M \nu ^2}\epsilon_{\mu \nu \lambda \sigma} 
      q^\lambda (p \cdot q s^\sigma - s \cdot q p^\sigma )
\notag \\
& 
     -b_1 r_{\mu \nu} 
     + \frac{1}{6} b_2 (s_{\mu \nu} +t_{\mu \nu} +u_{\mu \nu}) 
     + \frac{1}{2} b_3 (s_{\mu \nu} -u_{\mu \nu}) 
     + \frac{1}{2} b_4 (s_{\mu \nu} -t_{\mu \nu}) ,
\label{eqn:w-1}
\end{align}
by the eight structure functions including the new ones $b_{1-4}$.
Here, the kinematical coefficients $r_{\mu \nu}$, $s_{\mu \nu}$, 
$t_{\mu \nu}$, and $u_{\mu \nu}$ are defined as
\begin{align}
r_{\mu \nu} = & \frac{1}{\nu ^2}
   \bigg [ q \cdot E ^* (\lambda_f) q \cdot E (\lambda_i) 
           - \frac{1}{3} \nu ^2  \kappa \bigg ]
   \hat{g}_{\mu \nu}, 
\ \ \ \ 
s_{\mu \nu} =  \frac{2}{\nu ^2} 
   \bigg [ q \cdot E ^* (\lambda_f) q \cdot E (\lambda_i) 
           - \frac{1}{3} \nu ^2  \kappa \bigg ]
\frac{\hat{p}_\mu \hat{p}_\nu}{M \nu}, \notag \\
t_{\mu \nu} = & \frac{1}{2 \nu ^2}
   \bigg [ q \cdot E ^* (\lambda_f) 
           \left\{ \hat{p}_\mu \hat E_\nu (\lambda_i) 
                 + \hat{p} _\nu \hat E_\mu (\lambda_i) \right\}
   + \left\{ \hat{p}_\mu \hat E_\nu^* (\lambda_f)  
           + \hat{p}_\nu \hat E_\mu^* (\lambda_f) \right\}  
     q \cdot E (\lambda_i) 
   - \frac{4 \nu}{3 M}  \hat{p}_\mu \hat{p}_\nu \bigg ] ,
\notag \\
u_{\mu \nu} = & \frac{M}{\nu} 
   \bigg [ \hat E_\mu^* (\lambda_f) \hat E_\nu (\lambda_i) 
          +\hat E_\nu^* (\lambda_f) \hat E_\mu (\lambda_i) 
   +\frac{2}{3}  \hat{g}_{\mu \nu}
   -\frac{2}{3 M^2} \hat{p}_\mu \hat{p}_\nu \bigg ] .
\end{align}
\end{widetext}

In these equations, 
$k$ and $k'$ are initial and final lepton momenta,
$M$, $p$, and $q$ are hadron mass, hadron momentum,
and momentum transfer, $\nu$ and $Q^2$ are defined by $\nu ={p \cdot q}/{M}$,
and $Q^2=-q^2>0$, $p_{_{X}}$ is given by $p_{_{X}} =  p+q$,
and $\epsilon_{\mu \nu \lambda \sigma}$ is an antisymmetric 
tensor with the convention $\epsilon_{0123}=+1$.
The notations $\hat{g}_{\mu \nu}$ and $\hat{p}_\mu$ are used
for satisfying the current conservation 
$q^\mu W _{\mu \nu} = q^\nu W _{\mu \nu}=0$, and they are defined by
\begin{align}
\hat{g}_{\mu \nu} \equiv  g_{\mu \nu} -\frac{q_\mu q_\nu}{q^2}, \ \ 
\hat{a}_\mu \equiv a_\mu -\frac{a \cdot q}{q^2} q_\mu .
\label{eqn:hat}
\end{align}
Furthermore, $\kappa$ is defined by $\kappa= 1+{Q^2}/{\nu^2}$, 
$s^\mu$ is the spin vector of the spin-one hadron,
and $E^\mu$ is the polarization vector of the spin-one hadron 
with the constraints, $p \cdot E =0$ and $E^* \cdot E =-1$.
The polarization vector is taken as the spherical unit vectors:
\begin{align}
  E^\mu (\lambda= \pm 1) & = \frac{1}{\sqrt{2}}(0,\mp 1, -i,0), 
\nonumber \\
  E^\mu (\lambda=0) & = (0,0,0,1) ,
\end{align}
and its relation to the the spin vector is given by
\begin{equation}   
(s_{\lambda_f \lambda_i})^{\mu}
      = -\frac{i}{M} \epsilon ^{\mu \nu \alpha \beta} 
                E^*_\nu (\lambda_f) E_\alpha (\lambda_i) p_\beta .
\end{equation}
In these equations, we explicitly denoted the initial and 
final spin states by $\lambda_i$ and $\lambda_f$, respectively,
because off-diagonal terms with $\lambda_f \ne \lambda_i$ are 
generally needed to discuss higher-twist contributions \cite{hjm89}.

Among the four new structure functions $b_{1-4}$, $b_3$ and $b_4$ are
higher-twist functions, and the twist-two functions $b_1$ and $b_2$
are related with each other by the Callan-Gross like relation $2x b_1=b_2$
in the Bjorken scaling limit. The functions $b_1$ and $b_2$ are expressed
by the tensor-polarized parton distribution functions $\delta_T f (x)$
defined by \cite{delta-T-notation}
\begin{align}
\! \!
\delta_{_T} f (x,Q^2) \equiv f^0 (x,Q^2) 
          - \frac{f^{+1} (x,Q^2) +f^{-1} (x,Q^2)}{2},
\label{eqn:tensor-pdf}
\end{align}
where 
$f^\lambda$ indicates an unpolarized parton distribution
in the hadron spin state $\lambda$.
Using the tensor-polarized quark and antiquark distributions, 
we have the structure function 
\begin{align}
b_1 (x,Q^2) = \frac{1}{2} \sum_i e_i^2 
      \, \left [ \delta_{_T} q_i (x,Q^2) 
      + \delta_{_T} \bar q_i (x,Q^2)   \right ] , 
\label{eqn:b1-parton}
\end{align}
in the parton model. Here, $e_i$ is the charge of the quark flavor $i$.

There is sum rule for $b_1$ based on the parton model \cite{b1-sum,tensor-pdfs}
in the similar way to the Gottfried sum rule \cite{flavor3}:
\begin{align}
\! \! \!
\int dx \, b_1 (x) 
    & = - \lim_{t \to 0} \frac{5}{24} \, t \, F_Q (t) 
\nonumber \\
  & \ \ \ 
     + \frac{1}{9} \int dx
      \, \left [ \, 4 \, \delta_{_T} \bar u (x) +  \delta_{_T} \bar d (x) 
                     +   \delta_{_T} \bar s (x)  \, \right ] ,
\nonumber \\
 \int \frac{dx}{x}
 \, [F_2^p & (x) - F_2^n (x) ] 
   =  \frac{1}{3} 
   +\frac{2}{3} \int dx \, [ \bar u(x) - \bar d(x) ] ,
\label{eqn:b1-sum-gottfried}
\end{align}
where $F_Q(t)$ is the electric quadrupole form factor for
the spin-one hadron, and this first term vanishes:
$\lim_{t \to 0} \frac{5}{24} t F_Q (t)=0$.
These relations are derived in the parton model, and they are
not rigorous ones obtained, for example, by the current algebra
\cite{ioffe-book}. Therefore, depending on small-$x$ behavior 
of $\bar u -\bar d$ (or $F_2^p - F_2^n$) and $\delta_T \bar q$ (or $b_1$),
these sums may diverge although $\bar u$ and $\bar d$ are currently 
assumed to be equal at very small $x$ where experimental data do not exist. 
These sum rules should be considered as guidelines based on 
the parton model. In any case, as the violation of the Gottfried sum
rule created the field of flavor dependence in antiquark distributions
and studies of nonperturbative mechanisms behind it, the deviation
from zero for the $b_1$ sum could probe the interesting tensor-polarized
antiquark distributions $\delta_T \bar q$
and possibly exotic mechanisms behind them
as shown in Eq. (\ref{eqn:b1-sum-gottfried}).

The HERMES collaboration reported the first measurement on $b_1$
\cite{hermes05}, and its data were analyzed to obtain possible 
tensor-polarized PDFs in Ref. \cite{tensor-pdfs}.
Since there is little information for the tensor-polarized PDFs
at this stage, we need to introduce bold assumptions for 
extracting the distributions from the experimental data.
However, the $b_1$ seems to have a node at a medium $x$ point.
First, a conventional convolution description for $b_1$ indicates
an oscillatory $x$ dependence \cite{hjm89,kh91}. 
Namely, it is negative at $x>0.5$ and it turns into positive
at $x<0.5$. Second, the sign change is also suggested by the HERMES data
at $x = 0.2 - 0.3$. Third, the $b_1$ sum needs the node to satisfy 
$\int dx \, b_1 (x) =0$. From these ideas, we considered that 
tensor-polarized PDFs for the deuteron are expressed 
by the unpolarized PDFs multiplied by a function $\delta_{_T} w(x)$
at $Q_0^2$ as \cite{tensor-pdfs}
\begin{align}
\delta_{_T} q_v^D (x,Q_0^2) & \equiv \delta_{_T} u_v^D (x,Q_0^2) 
                         = \delta_{_T} d_v^D (x,Q_0^2) 
\nonumber \\
         & = \delta_{_T} w(x) \, \frac{u_v (x,Q_0^2)  +d_v (x,Q_0^2) }{2} , 
\nonumber \\
\delta_{_T} \bar q^D (x,Q_0^2)  & \equiv \delta_{_T} \bar u^D (x,Q_0^2) 
                            = \delta_{_T} \bar d^D (x,Q_0^2) 
\nonumber \\
                          & = \delta_{_T}      s^D (x,Q_0^2) 
                            = \delta_{_T} \bar s^D (x,Q_0^2)                     
\nonumber \\
   &   = \alpha_{\bar q} \, \delta_{_T} w(x) \,
\nonumber \\
   & \! \! \! \! \! \! \! \! \!  \! \! \! \! \! \! \! \! \! \! \! \! \! 
   \times
    \frac{2 \bar u(x,Q_0^2)  +2 \bar d(x,Q_0^2)  
    +s(x,Q_0^2)  + \bar s(x,Q_0^2) }{6} .
\label{eqn:dw(x)}
\end{align}
Here, $D$ indicates the deuteron and the capital letter is used
to avoid a confusion with a d-quark, and
the modification function $\delta_{_T} w(x)$
is expressed by a simple polynomial form with
a node at $x=x_0$:
\begin{equation}
\delta_{_T} w(x) = a x^b (1-x)^c (x_0-x) .
\label{eqn:dw(x)-abc}
\end{equation}
It means that a certain fraction of the unpolarized PDFs
is tensor-polarized, and this fraction is given by $\delta_{_T} w (x)$
or $\alpha_{\bar q} \, \delta_{_T} w (x)$ for valence-quark and antiquark
distributions, respectively. In the analysis of Ref. \cite{tensor-pdfs},
the scale is taken as the average $Q^2$ of the HERMES experiment 
($Q_0^2 = 2.5$ GeV$^2$).
The parameters $a$, $b$, $c$, and $x_0$ are determined by a $\chi^2$
analysis of the HERMES data with the leading-order expression 
for $b_1$ in Eq. (\ref{eqn:b1-parton}) by considering two options:
\begin{itemize}
\vspace{-0.15cm}
\item[(1)] Set 1: There is no tensor-polarized antiquark distribution 
                  at the scale $Q_0^2$ ($\alpha_{\bar q} = 0$).
\vspace{-0.15cm}
\item[(2)] Set 2: There could be finite tensor-polarized antiquark distributions
                  at the scale $Q_0^2$ ($\alpha_{\bar q}$ is a free parameter).
\end{itemize}
The determined parameter values are 
$a = 0.221 \pm 0.174$ ($0.378 \pm 0.212$),
$b = 0.648 \pm 0.342$ ($0.706 \pm 0.324$),
$\alpha_{\bar q} = 3.20 \pm 2.75$ (fixed$=0$),
$c = \text{fixed}=1$ ($1$), and
$x_0 = 0.221$ (0.229) in the set-2 (set-1).

\begin{figure}[b]
\begin{center}
\includegraphics[width=0.35\textwidth]{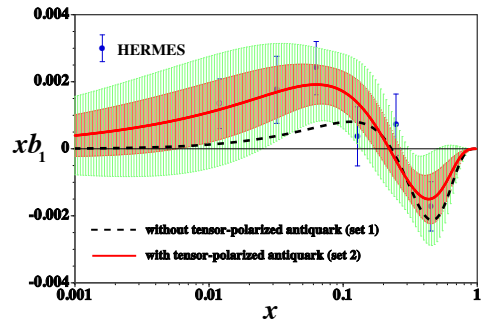}
\end{center}
\vspace{-0.5cm}
\caption{(Color online) $\chi^2$ analysis results are compared 
with HERMES data \cite{hermes05}.
The solid and dashed curves indicate theoretical results
with finite tensor-polarized antiquark distributions
(set-2, $\alpha_{\bar q} \ne 0$) and without them 
(set-1, $\alpha_{\bar q} = 0$), respectively. 
The open circle is the data at $Q^2<1$ GeV$^2$, 
and it is not included in the fit analysis \cite{tensor-pdfs}.
The uncertainties of the set-2 curve are shown by the bands
with $\Delta \chi^2=1$ and 3.53.}
\label{fig:xb1}
\end{figure}

Since the error matrix or Hessian is obtained in the $\chi^2$ analysis,
it is possible to show an error band for the obtained $b_1$ curve.
The details of the Hessian method is explained elsewhere \cite{hessian},
so that a brief outline is explained in the following.
Expanding $\chi^2$ around the minimum parameter set $\hat \xi$, 
we have the $\chi^2$ change ($\Delta\chi^2$) expressed by the 
second derivative matrix $H_{ij}$, which is called Hessian, as 
\begin{equation}
\Delta \chi^2 = \chi^2(\hat{\xi}+\delta \xi)-\chi^2(\hat{\xi})
=\sum_{i,j} H_{ij} \, \delta \xi_i \, \delta \xi_j .
\label{eq:hessian-chi2}
\end{equation}
Then, the error of a physics quantity $f(x)$ is calculated 
by the Hessian and  as
\begin{equation}
[\delta f(x)]^2=\Delta \chi^2 \sum_{i,j}
\left[ \frac{\partial f(x)}{\partial \xi_i}  \right]_{\hat \xi}
H_{ij}^{-1}
\left[ \frac{\partial f(x)}{\partial \xi_j}  \right]_{\hat \xi} .
\label{eq:erroe-M}
\end{equation}
In the set-2 analysis, there are three parameters 
$\xi_i = a$, $b$, and $\alpha_{\bar q}$. The node point $x_0$ 
is expressed by the parameter $b$. The derivatives with respect
to the parameters $\xi_i$ need to be calculated. 
As for the $\Delta \chi^2$ value, we may take $\Delta \chi^2 =1$. 
However, a larger value ($\Delta \chi^2 \sim N$, $N=$number of parameters) 
is often taken in the PDF analyses, and it is called a tolerance.
In Ref. \cite{hessian}, one-$\sigma$ range is given so that 
the confidence level becomes 68\% for the multi-parameter
normal distribution.
For $N=3$, it is $\Delta\chi^2 =3.53$.
In showing uncertainty bands, we take $\Delta\chi^2 = 1$ and 3.53
in this work. In any case, they are related with each other by
changing the band width by the scale $\sqrt{3.53}=1.88$.

Our optimum structure function $b_1$ of set-2 is shown in Fig. \ref{fig:xb1}
by the solid curve in comparison with the HERMES data. The set-1 result is 
shown by the dashed curve, which significantly deviates from the data
at small $x$ ($<0.1$). Although the analysis results depend on 
the parametrization function, the small-$x$ data cannot be explained
if there is no contribution from the tensor-polarized antiquark
distributions. The set-2 is a reasonable fit to the experimental measurements.
Two uncertainty bands are shown in Fig. \ref{fig:xb1} by taking
$\Delta \chi^2 =1$ and 3.53. The set-1 curve is in the error-band 
boundary at $x<0.1$ by considering the uncertainties, which indicates
that the set-1 function could be marginally consistent with
the measurements. It is because the errors of the HERMES measurements 
are large and hence the function $b_1$ still have large errors.

\begin{figure}[t!]
\vspace{0.3cm}
\begin{center}
\includegraphics[width=0.35\textwidth]{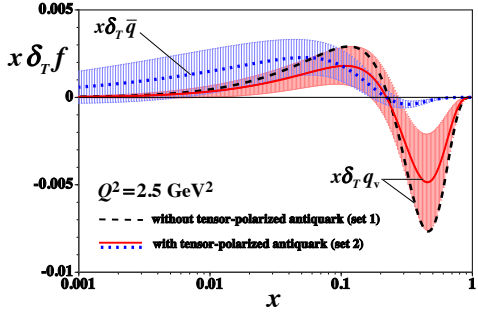}
\end{center}
\vspace{-0.5cm}
\caption{(Color online) Obtained tensor-polarized PDFs
at $Q^2 =2.5$ GeV$^2$.
The solid and dashed curves indicate the tensor-polarized
valence-quark distributions $x \delta_T q_v (x)$
with finite tensor-polarized antiquark distributions
(set-2, $\alpha_{\bar q} \ne 0$) and without them 
(set-1, $\alpha_{\bar q} = 0$), respectively. 
The dotted curve is the tensor-polarized antiquark distribution 
$x \delta_T \bar q (x)$ in the set-2 \cite{tensor-pdfs}.
The uncertainty bands of $\Delta \chi^2 =1$ are shown.}
\label{fig:tensor-pdfs}
\vspace{-0.3cm}
\end{figure}

Then, the determined tensor-polarized PDFs are shown in 
Fig. \ref{fig:tensor-pdfs}. The tensor-polarized valence-quark and 
antiquark distributions are shown for the set-2 by the solid and
dotted curves, and the valence-quark distribution of the set-1
is shown by the dashed curves. All the distributions are negative
at large $x$ ($>0.2$), and they turn into positive at the node point
around $x\sim 0.2$. These are distributions at $Q^2=2.5$ GeV$^2$.
Since these distributions are the only ones determined by
the $\chi^2$ analysis of existing $b_1$ data in a model
independent way, we use them for estimating
the tensor-polarization asymmetry in proton-deuteron Drell-Yan 
process at Fermilab.
However, one should note that the tensor-polaried PDFs are
not determined well at this stage as shown by the error 
bands in Fig. \ref{fig:tensor-pdfs}, and
future experimental progress is obviously needed for $b_1$.
Only the uncertainty bands of $\Delta \chi^2 =1$ are shown in 
Fig. \ref{fig:tensor-pdfs}. Hereafter, $\Delta \chi^2 =1$ bands
are shown in figures.

\subsection{Polarized proton-deuteron Drell-Yan process 
with tensor-polarized deuteron}
\label{pd-drell-yan}

Polarized proton-proton Drell-Yan processes have been theoretically
investigated extensively, and the studies are foundations 
for the RHIC (Relativistic Heavy Ion Collider)-spin project. 
However, polarized proton-deuteron processes have not 
been studied well partly because there was no actual experimental
project. In 1990's, a possible deuteron-beam polarization was
considered for RHIC \cite{rhic-d}; however, it was not realized.
On the other hand, the polarized proton-deuteron Drell-Yan
processes are possible with a polarized deuteron target,
and it is becoming a realistic project at Fermilab.

\begin{figure}[b]
   \vspace{-0.00cm}
\begin{center}
   \includegraphics[width=7.0cm]{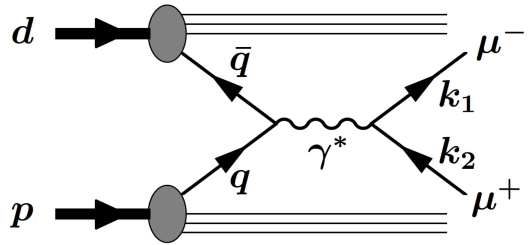}
\end{center} 
\vspace{-0.5cm}
\caption{Typical Drell-Yan process.}
\label{fig:drell-yan-qqbar}
\end{figure}

The formalisms of polarized proton-deuteron Drell-Yan processes
were investigated in Ref. \cite{pd-drell-yan}. 
The cross section for the Drell-Yan process $p + d \to \mu^+ \mu^- +X$ 
is given by a lepton tensor multiplied by the hadron tensor
\begin{align}
\! 
W_{\mu\nu}^{DY}  \! =  \! \frac{1}{4\pi M} \! 
   \int  \! d^{\,4} \xi \, e^{-i(k_1+k_2)\cdot \xi}
   \langle \, p \, d \, | \, J_\mu^{\,em} (\xi) J_\nu^{\,em}(0) \, 
   | \, p \, d \, \rangle  ,
\label{eqn:dy-hadron-tensor}
\\[-0.72cm] \nonumber
\end{align}
where $k_1$ and $k_2$ are momenta for $\mu^-$ and $\mu^+$, respectively.
The leading subprocess, which contributes to the cross section,
is the quark-antiquark annihilation process $q\bar q \to \mu^+ \mu^-$
shown in Fig. \ref{fig:drell-yan-qqbar}.

According to the general formalism for $W_{\mu\nu}^{DY}$ by using
Hermiticity, parity conservation, and time-reversal invariance
\cite{pd-drell-yan}, there exist 108 structure functions 
in the proton-deuteron (pd) Drell-Yan processes
instead of 48 functions in the proton-proton (pp) Drell-Yan. 
There are 60 new structure functions due to the spin-one nature 
of the deuteron. However, most of them are higher-twist functions 
and all of them are not important especially at the first stage.
Because it is too lengthy to write these structure functions, 
we explain only the essential functions associated with 
the leading-twist part of the deuteron tensor structure.

In the pp Drell-Yan processes, there are spin asymmetries 
by combinations of the unpolarized state (U)
together with longitudinal (L) and transverse (T) polarizations: 
$< \! \sigma \! >$, $A_{LL}$, $A_{TT}$, $A_{LT}$, and 
$A_{T} \, (\, =A_{UT}, \, A_{TU})$.
In the pd Drell-Yan, additional tensor-polarization asymmetries 
exist, and the following spin asymmetries could be investigated:
\begin{alignat}{8}
& \! \! \! \! \! \! \!
< \! \sigma \! >, \, & & 
A_{LL}, \,             & &
A_{TT}, \,             & &
A_{LT}, \,             & &
A_{TL}, \,             & &
A_{UT}, \,             & &
A_{TU}, \,             & &
\nonumber \\
& \! \! \! \! \! \! \! \,
A_{UQ_0}, \,         & &     
A_{TQ_0}, \,           & &
A_{UQ_1}, \,           & &
A_{LQ_1}, \,           & &
A_{TQ_1}, \,           & &
A_{UQ_2}, \,           & &
A_{LQ_2}, \,           & &
A_{TQ_2},
\end{alignat}
where $Q_0$, $Q_1$, and $Q_2$ indicate three tensor polarizations
depending on polarization direction \cite{pd-drell-yan,sk14}.
In particular, we investigate the tensor-polarization asymmetry
\begin{equation}
\!
A_{UQ_0} \! = \! \frac{1}{2 \! < \! \sigma \! >}     
         \bigg [ \sigma(\bullet , 0_L)
            - \frac{ \sigma(\bullet , +1_L) 
                    +\sigma(\bullet , -1_L) }{2} \bigg ] , \! \! 
\label{eqn:a-uq0}
\end{equation}
in this work. Here, $\bullet$ indicates the unpolarized proton.

\begin{figure}[b]
   \vspace{-0.00cm}
\begin{center}
   \includegraphics[width=6.5cm]{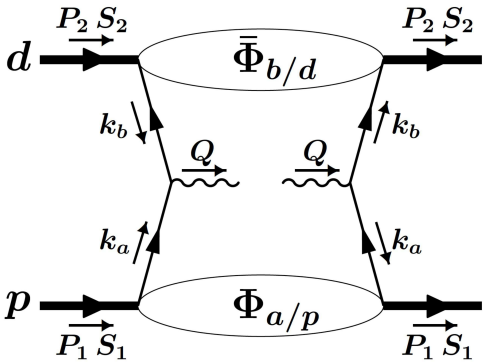}
\end{center} 
\vspace{-0.5cm}
\caption{Parton model for Drell-Yan process.}
\label{fig:drell-yan-parton}
\end{figure}

The Drell-Yan cross sections and spin asymmetries can be expressed
in terms of PDFs. As shown in Fig. \ref{fig:drell-yan-parton},
the leading contribution to the hadron tensor is 
generally given by quark and antiquark correlation functions as
\begin{align}
& W_{\mu \nu}^{\, q\bar q} = \frac{1}{3} \sum_{a, b} \delta_{b \bar{a}} \, e_a^2 
            \int d^4 k_a \, d^4 k_b \, \delta^4 (k_a + k_b - Q) \, 
\nonumber \\
   & \ \ \ \ \ \ \ 
   \times  \text{Tr} \left [ \, \Phi_{a/p} (P_1 S_1; k_a) \gamma_{\mu} 
           \bar{\Phi}_{b/d} (P_2 S_2; k_b) \gamma_{\nu} \, \right ] ,
\label{eqn:w-2}
\end{align}
where $k_a$ and $k_b=k_{\bar a}$ are quark and antiquark momenta,
the color average and summations are taken by $3 \cdot (1/3)^2$, and
the correlation functions $\Phi_{a/p}$ and $\bar \Phi_{\bar a/d}$ are
defined by the quark field $\psi$ as
\begin{align}
\!
\left ( \Phi_{a/p} \right )_{ij} & =
         \! \! \int \! \frac{d^4 \xi}{(2 \pi)^4} \, e^{i k_a \cdot \xi}
\, \langle \, P_1 S_1 \, | \, 
\bar{\psi}_j^{(a)}(0) \, \psi_i^{(a)}(\xi) \, | \, P_1 S_1
\, \rangle ,
\nonumber \\
\!
\left ( \bar{\Phi}_{\bar a/d} \right )_{ij} & =
          \! \! \int \! \frac{d^4 \xi}{(2 \pi)^4} \, e^{i k_{\bar a} \cdot \xi}
\, \langle \, P_2 S_2 \, | \, \psi_i^{(a)}(0) \, \bar{\psi}_j^{(a)}(\xi) \,
 | \, P_2 S_2
\, \rangle .
\end{align}
Here, link operators for satisfying the gauge invariance 
are not explicitly written.
The correlation functions are expressed by unpolarized,
longitudinally-polarized, transversity distributions,
and new tensor-polarized distributions.
There is another contribution obtained by exchanging quark and antiquark
($W_{\mu \nu}^{\, \bar q q}$).

Now, we express the Drell-Yan cross section, structure functions,
and the tensor-polarization asymmetry in the parton model.
There are 19 structure functions for the proton-deuteron Drell-Yan
in the parton model; however, the number becomes
four after the integration over the transverse momentum of $Q$.
Denoting $\overline F = \int d^2 Q_T F$ for the structure function $F$,
we have the pd Drell-Yan cross section \cite{pd-drell-yan}:
\begin{align}
\frac{d \sigma}{dx_1 \, dx_2 \, d \Omega} & = \frac{\alpha^2}{4 \, Q^2} \,
      \bigg [ \,  (1 + \cos^2 \theta) \bigg\{ \, 
            \overline W_T 
            + \frac{1}{4} \lambda_1 \lambda_2 \, \overline V_T^{\, LL}
\nonumber \\
      & \ \ \ \ \ \ \ \ \ \ \ 
           + \frac{2}{3}  \left( 2 \, |\vec S_{1T}|^2 - \lambda_1^2 \right)
                               \, \overline V_T^{\, UQ_0} \, \bigg\}
\nonumber \\
&  \! \! \! \! \! \! \! \! \! \! \! \! \! \! \! \! \!
+  \sin^2 \theta \, |\vec S_{1T}| \, |\vec S_{2T}|
             \cos(2\phi-\phi_1-\phi_2) \, \overline U_{2,2}^{\, TT}
        \, \bigg ] ,
\end{align}
where $\theta$ and $\phi$ are polar and azimuthal angles of 
the vector $\vec k_1 - \vec k_2$ for the dimuon,
$\lambda_1$ and $\lambda_2$ are proton and deuteron helicities,
and $\vec S_{1T}$ and $\vec S_{2T}$ are the transverse spin vectors
for the proton and deuteron defined by the angle $\phi_1$ and 
$\phi_2$ as
$\vec S_{1T} = |\vec S_{1T}| \, (\cos\phi_1,\, \sin\phi_1,\, 0)$,
$\vec S_{2T} = |\vec S_{2T}| \, (\cos\phi_2,\, \sin\phi_2,\, 0)$.
The variables $x_1$ and $x_2$ are lightcone momentum fractions 
for the partons in the proton and deuteron, respectively.
Neglecting hadron masses and transverse momentum of the dimuon, 
we have the relation $Q^2 = M_{\mu\mu}^{2} = x_1 x_2 s$ 
\cite{field-book}. The rapidity of the muon pair is given by
$y =(1/2) \ln ((E_{\mu\mu}+P_{\mu\mu,L})/(E_{\mu\mu}-P_{\mu\mu,L}))
   =(1/2) \ln (x_1 /x_2)$, where $E_{\mu\mu}$ and $P_{\mu\mu,L}$
are dimuon energy and longitudinal momentum. The momentum fractions
$x_1$ and $x_2$ are expressed by these external variables as
$x_1 = \sqrt{\tau} e^{y}$ and $x_2 = \sqrt{\tau} e^{-y}$.
The dimuon transverse momentum is generally small in comparison with
the dimuon mass in the Fermilab experiment \cite{Fermilab-MI}.
In the parton model, the structure functions are expressed 
by the parton distributions
for the process $q+\bar q \rightarrow \mu^+ + \mu^-$ as
\begin{align}
\overline W_T     & = \frac{1}{3} \sum_i e_i^2 \, 
                 q_i (x_1,Q^2) \, \bar q_i (x_2,Q^2)
                 + (q \leftrightarrow \bar q) , 
\nonumber \\
\overline V_T^{\, LL} & = - \frac{4}{3} \sum_i e_i^2 \, 
          \Delta q_i (x_1,Q^2) \, \Delta \bar q_i (x_2,Q^2) 
                 + (q \leftrightarrow \bar q) , 
\nonumber \\
\overline U_{2,2}^{\, TT} & = \frac{1}{3} \sum_i e_i^2 \, 
          \Delta_{_T} q_i (x_1,Q^2) \, \Delta_{_T} \bar q_i (x_2,Q^2)  
                 + (q \leftrightarrow \bar q) , 
\nonumber \\ 
\! \!
\overline V_T^{\, UQ_0} & 
            = \frac{1}{6} \sum_i e_i^2 \, 
                       q_i (x_1,Q^2) \, \delta_{_T} \bar q_i (x_2,Q^2) 
                 + (q \leftrightarrow \bar q) , 
\end{align}
where $\Delta q_i$ and $\Delta_{_T} q_i$ are longitudinally-polarized
and transversity distributions. 
The terms $(q \leftrightarrow \bar q)$ indicate 
the contributions from $\bar q$(in p)+$q$(in d)$\rightarrow \mu^+ + \mu^-$
by the replacements $q \leftrightarrow \bar q$ in the first terms.

In this article, we are interested in investigating the tensor-polarization
asymmetry $A_{UQ_0}$ of Eq. (\ref{eqn:a-uq0}), and it is expressed 
in the parton model as \cite{pd-drell-yan,sk14}
\begin{align}
& A_Q   \equiv 2 A_{UQ_0} 
         = \frac{2 \, \overline V_T^{\, UQ_0}}{\overline W_T}
\nonumber \\
 & \! \! \! \! \! \!
    = \frac{\sum_i e_i^2 \, 
            \left[ \, q_i(x_1,Q^2) \, \delta_{_T} \bar q_i(x_2,Q^2)
                + \bar q_i(x_1,Q^2) \, \delta_{_T} q_i(x_2,Q^2) \, \right] }
                {\sum_i e_i^2 \, 
            \left[ \, q_i(x_1,Q^2) \, \bar q_i(x_2,Q^2)
                + \bar q_a(x_1,Q^2) \, q_i(x_2,Q^2) \, \right] } ,
\label{eqn:a-uq0-parton}
\end{align}
where the factor of 2 is multiplied to define $A_Q$ \cite{delta-T-notation}
so that the asymmetry becomes the simple ratio of the unpolarized PDFs
$f (x)$ and the tensor-polarized PDFs $\delta_T f (x)$,
although there is nothing wrong in the formalism of Ref. \cite{pd-drell-yan}.
The scale $Q^2$ is given by the dimuon mass
$Q^2 = M_{\mu\mu}^{\, 2} =(k_1+k_2)^2= (k_a+k_b)^2 = x_1 \, x_2 \, s$,
the unpolarized PDFs $f (x,Q^2)$ and tensor-polarized PDFs 
$\delta_T f (x,Q^2)$ should be obtained at this $Q^2$ scale 
for estimating the asymmetry $A_Q$.
The unpolarized PDFs are well known, and we could use the available 
parametrization for the tensor-polarized PDFs in Eq. (\ref{eqn:dw(x)}) and 
Fig. \ref{fig:tensor-pdfs} for our numerical calculations of $A_Q$.

\section{Results} 
\label{results}

Because the tensor-polarized PDFs are obtained at $Q^2=2.5$ GeV$^2$
in Eq. (\ref{eqn:dw(x)}), they need to be evolved to the $Q^2$ points 
of the Fermilab Drell-Yan experiment. The proton beam energy is 
$E_p=120$ GeV for the Fermilab Main Injector (MI), so that the center-of-mass 
energy squared is $s = (p_1 + p_2)^2 = M_p^2 + M_d^2 + 2 M_d E_p$ 
in the proton-deuteron Drell-Yan process with the proton (deuteron) 
four momentum $p_1$ ($p_2$) and its mass $M_p$ ($M_d$). 
Then, the dimuon mass $M_{\mu\mu}$, which is
equal to $Q^2$, is given by $Q^2=M_{\mu\mu}^2 = x_1 x_2 s$.
So far, the dimuon-mass region of $4^2$ GeV$^2$$< M_{\mu\mu}^2 < 9^2$ GeV$^2$
is measured between the $J/\psi$ and $\Upsilon$ resonances.
Therefore, the $Q^2$ evolution of the tensor-polarized PDFs are necessary
for estimating the tensor-polarization asymmetry for 
the Fermilab experiment.

\subsection{Scale dependence of the tensor-polarized \\
parton distribution functions} 
\label{scale-dependence}

The $Q^2$ evolution of $b_1$ and the tensor-polarized PDFs is
rarely discussed, so that we briefly mention it here 
along the explanation of Ref. \cite{hjm89}.
The operator product expansion (OPE) was studied within the twist-two level,
and the time-ordered product of two electromagnetic currents is expressed 
by twist-two vector and axial-vector operators,
$O_V^{\,\mu_1 \cdots \mu_n}$ and $O_A^{\,\mu_1 \cdots \mu_n}$. 
The vector operators are defined by
the quark field $\psi$ and the covariant derivative $D^{\,\mu}$ as
\begin{align}
\!
O_V^{\,\mu_1 \cdots \mu_n} = \frac{1}{2} \left ( \frac{i}{2} \right )^{n-1}
    \! \! \! 
    S \left [ \, \overline \psi \, \gamma^{\,\mu_1} 
       \overset{\text{\scriptsize$\leftrightarrow$}}{D}\,^{\mu_2} \cdots
       \overset{\text{\scriptsize$\leftrightarrow$}}{D}\,^{\mu_n} 
       \, e_q^{\,2}  \, \psi \, \right ]  ,
\label{eqn:V-operators}
\end{align}
where $S$ indicates the symmetrization of the indices ($\mu_1 \cdots \mu_n$)
and removal of the trace. The structure functions $F_{1,2}$ and $b_{1,2}$ 
can be obtained from the vector operators $O_V^{\,\mu_1 \cdots \mu_n}$, 
whose matrix elements are expressed as
\begin{align}
& \left. \langle \, p, E \, \right | \, O_V^{\mu_1 \cdots \mu_n} \, 
  \left | \, p, E \, \rangle  \right.
= S  \bigg [ \, a_n p^{\,\mu_1} \cdots p^{\,\mu_n} 
\nonumber \\
& \ \ \ \ \ \ \ 
+d_n \left ( E^{* \mu_1} E^{\,\mu_2} 
                      - \frac{1}{3} \, p^{\,\mu_1} p^{\,\mu_2} \right )
                      \, p^{\,\mu_3} \cdots p^{\,\mu_n}
    \, \bigg ] .
\label{eqn:OV-matrix-element}
\end{align}
The coefficients $a_n$ and $d_n$ are associated with the structure
functions $F_1$ and $b_1$, respectively.
Therefore, as explained in Ref. \cite{hjm89}, the $b_1$ and $b_2$ 
are obtained from the same vector operators $O_V^{\,\mu_1 \cdots \mu_n}$.
Then, their anomalous dimensions and also coefficient functions
in the OPE are common in $F_{1,2}$ and $b_{1,2}$, so that 
the $Q^2$ evolutions of $b_{1,2}$ are the same as the ones for $F_{1,2}$.
Namely, the $Q^2$ evolution of the tensor-polarized PDFs is calculated
by the same DGLAP (Dokshitzer-Gribov-Lipatov-Altarelli-Parisi)
evolution equations with the replacement of the unpolarized PDFs by
the tensor-polarized PDFs. Therefore, the DGLAP $Q^2$-evolution 
code, for example in Ref. \cite{bf1}, can be used for calculating
$Q^2$ variations of the tensor-polarized PDFs.

\begin{figure}[t!]
\begin{center}
\includegraphics[width=0.42\textwidth]{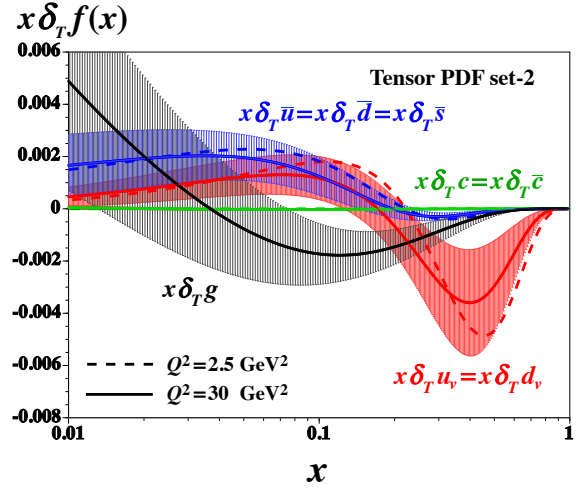}
\end{center}
\vspace{-0.6cm}
\caption{(Color online) $Q^2$ evolution results 
of set-2 tensor-polarized PDFs are shown. The dashed curves 
are the initial distributions at $Q^2 =2.5$ GeV$^2$,
and the solid curves indicate the tensor-polarized PDFs 
at $Q^2=30$ GeV$^2$.
}
\label{fig:tensor-evolution-2}
\end{figure}

\begin{figure}[t!]
\begin{center}
\includegraphics[width=0.42\textwidth]{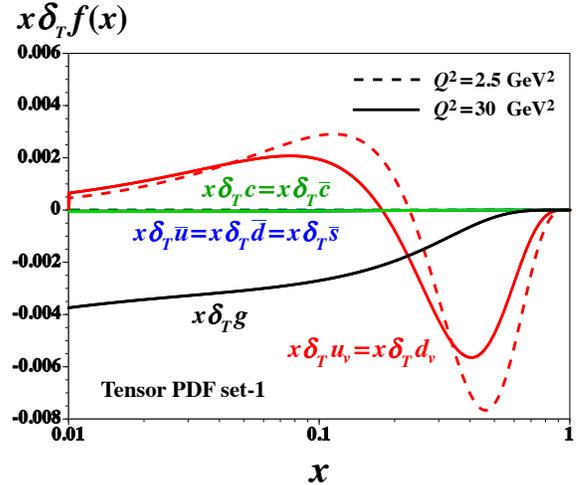}
\end{center}
\vspace{-0.6cm}
\caption{(Color online) $Q^2$ evolution results 
of set-1 tensor-polarized PDFs are shown. The dashed curves 
are the initial distributions at $Q^2 =2.5$ GeV$^2$,
and the solid curves indicate the tensor-polarized PDFs 
at $Q^2=30$ GeV$^2$.}
\label{fig:tensor-evolution-1}
 \vspace{-0.2cm}
\end{figure}

The tensor-polarized PDFs in Eq. (\ref{eqn:dw(x)}) and
Fig. \ref{fig:tensor-pdfs} at $Q^2=2.5$ GeV$^2$
are evolved to the $Q^2$ points, $Q^2=M_{\mu\mu}^2 = x_1 x_2 s$,
for the Fermilab-MI kinematics by the standard DGLAP evolution equations
\cite{bf1}. At the initial scale of $Q_0^2=2.5$ GeV$^2$, 
the tensor-polarized charm and gluon distributions are assumed to be zero:
$\delta_T c (x,Q_0^2)=\delta_T \bar c (x,Q_0^2)=\delta_T g (x,Q_0^2)=0$.
The uncertainties of the evolved PDFs are shown by the bands
for $\Delta \chi^2 =1$, and 
the uncertainty of the gluon distribution is also estimated.
The evolution results of the set-2 tensor-polarized PDFs
are shown in Fig. \ref{fig:tensor-evolution-2}
from the initial $Q^2=2.5$ GeV$^2$ to $Q^2=30$ GeV$^2$, which
roughly corresponds to a typical $Q^2$ value of the Fermilab experiment.
Within this $Q^2$ variation, the PDFs do not change significantly
although the node position moves from $x = 0.22$ to $0.16$.
However, it is interesting to find a significant tensor-polarized
gluon distribution $\delta_T g (x)$ due to the $Q^2$ evolution 
although it is zero in the initial scale, and its $x$ dependence
is much different from the quark and antiquark distributions.
We use the evolution results at $Q^2= x_1 x_2 s$ for given
$x_1$ and $x_2$ for calculating the tensor-polarized spin
asymmetries at Fermilab in the next subsection.
For comparison, the evolution results of the set-1 tensor-polarized PDFs
are shown in Fig. \ref{fig:tensor-evolution-1}.
Here, there is no antiquark tensor-polarization at the initial
scale $Q^2=2.5$ GeV$^2$. Finite tensor-polarized antiquark 
distributions are obtained at $Q^2=30$ GeV$^2$ due to the $Q^2$ evolution;
however, they are still tiny as shown in the figure.
Because the Drell-Yan cross section is sensitive to the antiquark 
distributions, it leads to small tensor-polarization asymmetries.
At this stage, the set-2 distributions are more realistic ones
because they can explain the HERMES measurements.
It is also interesting to find a significant tensor-polarized
gluon distribution although the antiquark distributions are
very small in the set-1 parametrization.

\subsection{Tensor-polarization asymmetry \\ in proton-deuteron
Drell-Yan process} 
\label{fermilab-pd-drell-yan}
\vspace{-0.2cm}

\begin{figure}[b]
\begin{center}
\includegraphics[width=0.42\textwidth]{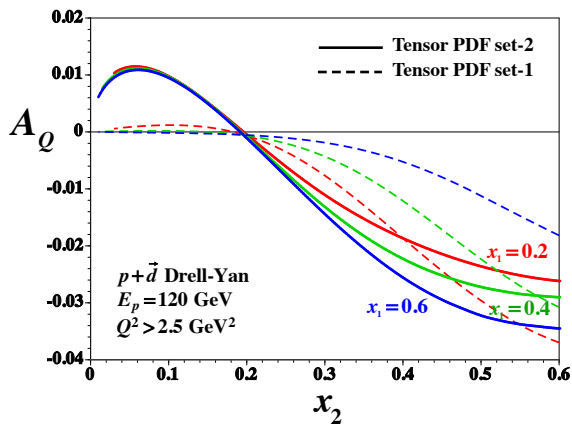}
\end{center}
\vspace{-0.6cm}
\caption{(Color online) Tensor-polarization asymmetries $A_Q$
are shown for $x_1=0.2$, 0.4 and 0.6 by using two sets of tensor
polarized PDFs. The dashed curves are for the set-1 without
the tensor-polarized antiquark distributions at the initial scale 
($Q^2=2.5$ GeV$^2$), the solid curves indicate the set-2 results.}
\label{fig:tensor-AQ}
\end{figure}

We show the obtained tensor-polarization asymmetries $A_Q$
in Fig. \ref{fig:tensor-AQ} at $x_1=0.2$, 0.4, and 0.6.
There are two-sets of tensor-polarized PDFs as shown
in Eq. (\ref{eqn:dw(x)}) and Fig. \ref{fig:tensor-pdfs}.
The tensor-polarized PDFs are evolved from $Q^2=2.5$ GeV$^2$ 
to the scale $Q^2 = M_{\mu\mu}^{\, 2}= x_1 x_2 s$
by taking the proton energy $E_p=120$ GeV with the fixed-target
deuteron for the Fermilab-MI experiment.
Since the MSTW2008 unpolarized LO (leading order) PDFs were used in
the $b_1$ analysis \cite{tensor-pdfs}, 
the MSTW2008-LO code \cite{MSTW2008} is
used for calculating the unpolarized PDFs at the same $Q^2$ points.
There are small nuclear corrections, usually within a few percent, 
in the unpolarized PDFs for the deuteron \cite{hkn07}, 
but they are neglected in this work.
In showing the curves, the experimental kinematical cut 
$4^2 < M_{\mu\mu}^{\, 2} < 9^2$ GeV$^2$ is not applied
except for the condition $Q^2=M_{\mu\mu}^{\, 2}>2.5$ GeV$^2$.
If $x_1$ is large, the large $x_2$ region cannot be reached
in the Fermilab experiment because of the condition
$4^2 < M_{\mu\mu}^{\, 2} < 9^2$ GeV$^2$.

We find in Fig. \ref{fig:tensor-AQ} that the tensor-polarization
asymmetry $A_Q$ is generally of the order of a few percent.
The set-1 asymmetries are rather small because
the antiquark tensor polarization does not exist at $Q^2=2.5$ GeV$^2$,
and it appears only by the $Q^2$ evolution. The set-2 asymmetries are
generally much larger. We believe that the set-2 results are more
reliable at this stage because they can explain the HERMES measurements 
including the small-$x$ region as shown in Fig. \ref{fig:xb1}. 
Since the HERMES $b_1$ data are taken in the region $0.012<x<0.452$,
our predictions should be reasonable ones for the symmetries 
in the Fermilab experiment, where the kinematical region 
$0.1<x_2<0.5$ will be probed.
There are large differences in asymmetries between set-1 and set-2.
However, in other words, it indicates the importance
to measure tensor-polarization asymmetry because it is the advantage 
of the Drell-Yan experiment to probe the antiquark distributions. 
Shadowing effects are effectively included in the HERMES data
at small $x$, possibly at $x<0.05$, our predictions include
such effects in the set-2. However, the Fermilab experiment
is not sensitive to the small $x_2$ region ($x_2 <0.05$),
so that we may wait for the electron-on collider project
\cite{eic} for small-$x$ measurements to probe shadowing effects
on $b_1$.

\begin{figure}[b]
\begin{center}
\includegraphics[width=0.42\textwidth]{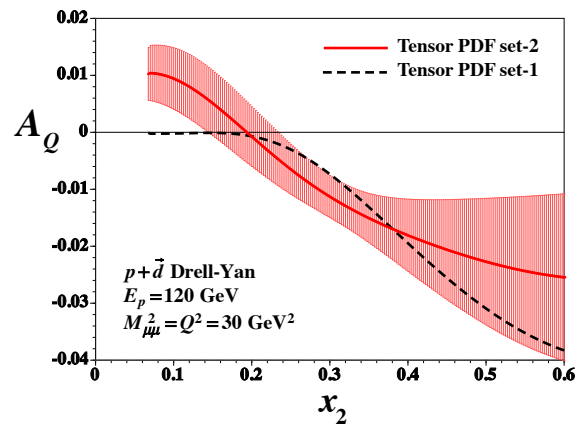}
\end{center}
\vspace{-0.6cm}
\caption{(Color online) 
Tensor-polarization asymmetries $A_Q$
are shown at the fixed $M_{\mu\mu}^{\, 2}=Q^2=30$ GeV$^2$
by using two sets of tensor polarized PDFs.
The dashed curve is for the set-1 without
the tensor-polarized antiquark distributions at the initial scale 
($Q^2=2.5$ GeV$^2$), the solid curve indicates the set-2 results.
The uncertainty band is shown for the set-2 curve.
}
\label{fig:tensor-AQ-error}
\end{figure}

One thing we need to check is the uncertainty of the tensor-polarized
PDFs in showing the asymmetries 
because they are not well determined as shown in 
Figs.\,\ref{fig:tensor-pdfs} and \ref{fig:tensor-evolution-2}.
We show the uncertainties for the set-2 at fixed $M_{\mu\mu}^2=Q^2=30$ GeV$^2$ 
in Fig.\,\ref{fig:tensor-AQ-error}. There are large differences 
in the asymmetries between set-1 and set-2; however, they are mostly
within the error bands, whereas the set-1 curve at small $x_2 (<0.2)$ 
is outside of the $\Delta \chi^2=1$ band. This small-$x_2$ discrepancy
is caused by the difference in handing the antiquark distributions
at $x<0.2$ originally in Fig.\,\ref{fig:tensor-pdfs}.
Although the error bands are large, we predict a finite tensor-polarized
asymmetry which could be investigated at Fermilab or other hadron
facilities.

In Eq. (\ref{eqn:b1-sum-gottfried}), 
we mentioned that tensor-polarized antiquark distributions 
are important to be measured for finding a possible mechanism
on the tensor structure in quark and gluon degrees of freedom.
Therefore, the Drell-Yan measurement is a valuable experiment 
which is very likely to create a new field of hadron physics. 
It is complementary to the JLab $b_1$ experiment which will start 
in a few years. There is also a plan to measure 
the tensor polarization $A_{zz}$ 
at JLab in the large-$x$ region \cite{azz},
and $b_1$ could be measured at EIC \cite{eic}.

Historically, the Fermilab Drell-Yan experiment played a crucial
role in establishing the flavor asymmetric antiquark distributions
$\bar u \ne \bar d$ \cite{flavor3}. This flavor asymmetry was 
suggested in the NMC (New Muon Collaboration) experiment 
by the violation of the Gottfried sum rule;
however, it was not obvious whether the NMC experiment could be
interpreted by a small-$x$ contribution without the flavor asymmetric
distributions. This issue was clarified by the Fermilab 
Drell-Yan experiment on the cross section ratio 
$\sigma_{pd}^{DY} / (2\sigma_{pp}^{DY})$, which directly probed
$\bar u \ne \bar d$.
In the same way, the tensor-polarized Drell-Yan experiment
should be valuable for probing the antiquark tensor polarization
directly. Hopefully, such an experiment will be done at Fermilab. 
In addition, it could be done at any facilities with high-energy 
hadron beams such as BNL (Brookhaven National Laboratory)-RHIC, CERN-COMPASS, 
J-PARC (Japan Proton Accelerator Research Complex) \cite{j-parc},
GSI-FAIR (Gesellschaft f\"ur Schwerionenforschung -Facility for 
Antiproton and Ion Research), and IHEP (Institute for High Energy Physics)
in Russia.

\vspace{-0.2cm}
\section{Summary}\label{summary}
\vspace{-0.2cm}

There exist new polarized structure functions for spin-one hadrons such
as the deuteron. In particular, the twist-two structure functions 
$b_1$ and $b_2$ of charged-lepton DIS are expressed in terms of
tensor-polarized PDFs. These functions could probe peculiar nature 
of hadrons in the sense that they should vanish if the internal 
constituents are in the S-wave and that HERMES $b_1$ measurements 
are much larger than the conventional deuteron-model estimate. 

New accurate measurements are planned at JLab by the electron DIS with
the tensor-polarized deuteron. Furthermore, a Fermilab Drell-Yan
experiment is now under consideration with the fixed tensor-polarized
deuteron target. For pursuing this experiment and allocating beam time
at Fermilab, it is crucial to estimate the magnitude of a possible 
tensor-polarization asymmetry theoretically. Using the optimum 
tensor-polarized PDFs obtained by analyzing the HERMES data, 
we estimated the tensor-polarization asymmetry by considering 
the Fermilab kinematics. We found that the asymmetry $A_Q$ is 
of the order of a few percent. 

It is a small quantity; however, we believe that it is worth for 
the measurement to find the physics mechanisms of tensor polarization
in the parton level. Especially, the Drell-Yan experiment should provide
important information on the tensor-polarized antiquark distributions.
It could lead to a new field of high-energy spin physics to probe
an exotic aspect in hadrons. Furthermore, we showed in our analysis
that a finite tensor-polarized gluon distribution should exist,
and it has never been studied experimentally. It is also 
an interesting future topic.

\begin{acknowledgements}
\vspace{-0.2cm}

The authors thank X. Jiang, D. Keller, A. Klein, and K. Nakano
for communications on a possible Fermilab Drell-Yan experiment with 
the tensor-polarized deuteron and R. L. Jaffe on scale dependence 
of the tensor-polarized PDFs and $b_1$.
This work was supported by Ministry of Education, Culture, Sports, 
Science and Technology (MEXT) KAKENHI Grant No. 25105010.
Q.-T.S is supported by the MEXT Scholarship for foreign students 
through the Graduate University for Advanced Studies.

\end{acknowledgements}




\begin{thebibliography}{00}
\bibitem{delta-T-notation} The overall factor 1/2 is introduced in $b_1$
      as it is used in $F_1$ and $g_1$ in expressing them by the unpolarized
      or longitudinally-polarized PDFs.
\end{thebibliography}
\end{document}